# SAGE-Net: Semantics-Augmented Geometric Encoder for Material Property Prediction

Guanghui Zhang[1], Yuxuan Yao[2], Kieran B. Spooner[1], Jun Yin[2], Dan Han[1*], David O. Scanlon[3], Lijun Zhang[1*]

[1] *State Key Laboratory of Integrated Optoelectronics, Key Laboratory of Automobile Materials of MOE, and School of Materials Science and Engineering, Jilin University, Changchun, China*

[2]*Department of Physics and Materials, The Hong Kong Polytechnic University, Kowloon, Hong Kong, China*

[3]*School of Chemistry, University of Birmingham, Edgbaston, Birmingham, B15 2TT, United Kingdom*

*Corresponding authors: hand@jlu.edu.cn, lijun_zhang@jlu.edu.cn

## Abstract:

Reliable structure–property modeling is crucial for accelerating materials discovery, where crystal graphs and structure-derived crystallographic descriptions provide complementary geometric and semantic information. Existing multimodal materials models primarily incorporate textual information through post-encoding fusion, latent-space alignment, or attention-based representation interaction mechanisms. However, in most cases, crystallographic semantics are introduced after structural encoding and therefore cannot directly guide the formation of atom-level crystal-graph representations. Here, we present Semantics-Augmented Geometric Encoder Network (SAGE-Net), a flexible multimodal framework that injects description-derived chemical and crystallographic semantics into geometric message passing. SAGE-Net introduces Semantic-Guided Message Passing (SGMP), which gates atom-level updates and enables crystallographic semantics to directly modulate local geometric interactions across multiple graph neural network (GNN) backbones. Across benchmarks covering bandgap, mechanical, transport-related properties, and synthesizability assessment, the SAGE-Net instantiated with different GNN backbones achieves the lowest MAE on eight out of ten JARVIS-DFT regression targets and delivers strong or highly competitive performance against both structure-based and multimodal baselines. For synthesizability assessment, the SAGE-Net demonstrate outstanding classification performance and high recall rates. Interpretability analysis unravels that SAGE-Net effectively captures physically interpretable crystallographic features, viz. space group, dimensionality, polyhedral environments, among others. Together, these results demonstrate SGMP-based SAGE-Net as a general and transferable framework for deeply integrated multimodal materials learning.

## INTRODUCTION

Accelerating the discovery of functional materials is a central goal in modern materials science and chemistry.[1-7] Materials design relies on establishing accurate structure–property relationships that enable the prediction of physicochemical behavior from atomic arrangements.[8-10] While first-principles methods such as density functional theory (DFT) serve as a theoretical benchmark, the computational cost for accurate predictions with high-level theory limits applicability for large-scale screening across complex systems.[11-13] Consequently, data-driven approaches, particularly machine learning (ML),[14] have emerged as powerful alternatives for rapid property prediction with increasing accuracy.[15-18]

A persistent bottleneck is that many chemically meaningful factors are not fully captured by atomic coordinates alone, but can be expressed in structure-derived crystallographic descriptions. These factors include oxidation states, coordination environments, and defect chemistry. Standard structure-based models typically do not incorporate such information explicitly.[12,19] Geometric deep learning (GDL) models such as Crystal Graph Convolutional Neural Networks (CGCNN), Atomistic Line Graph Neural Network (ALIGNN), Deeper Graph Neural Networks (DenseGNN) and Cartesian encoding graph neural network (CartNet) achieve strong performance by representing crystals as graphs, with atoms as nodes and interactions as edges.[20-24] However, standardized crystallographic descriptions can provide chemical and structural context that is complementary to crystal graph features.[12,19,25,26] In practice, structure-derived descriptions encode chemically relevant cues at multiple levels. They capture local chemistry such as oxidation states, bonding descriptors, and coordination environments, as well as recurring structural motifs such as polyhedra, corner- vs edge-sharing connectivity, and dimensionality, and crystallographic symmetry such as space group and crystal system.[27] Neglecting this semantic layer would reduce property-prediction accuracy.[19,25,26,28] It also hinders the discrimination between theoretically plausible structures and experimentally synthesizable candidates.

To address this challenge, multimodal learning has emerged as a strategy to enrich materials representations by coupling crystal structures with textual descriptions.[12,19,26,28] Existing multimodal materials models differ in how they couple structural and textual information. Crystal Multi-Modal Network (CrysMMNet) and Hybrid Large Language Model–Graph Neural Network (Hybrid-LLM-GNN) mainly follow a post-encoding fusion paradigm, where crystal-graph and text-derived representations are independently encoded, and then integrated at the prediction stage.[19,29] MultiMat further promotes cross-modal consistency through self-supervised alignment in a shared latent space, while MatMMFuse enhances representation-level interaction through attention-based fusion.[12,26] Despite these advances, the geometric message-passing backbone in these methods is generally constructed before, or largely independently of, crystallographic descriptions. As a result, textual cues such as symmetry, dimensionality, coordination motifs, and site environments do not directly condition atom-level feature updates during crystal-graph representation learning.

Motivated by this gap, we introduce SAGE-Net, a multimodal framework that injects crystallographic description-derived semantics into geometric message passing. Its core mechanism is Semantic-Guided Message Passing (SGMP), which conditions atom-level feature updates on textual crystallographic descriptors and addresses the decoupling inherent in late-fusion architectures. In this work, we show that interaction-based structure–description coupling improves both property prediction and materials screening. SAGE-Net is designed as a flexible framework that is not limited to a specific GNN backbone and can be instantiated with multiple structural encoders, including CGCNN, ALIGNN, DenseGNN-Lite, and CartNet. Across ten JARVIS-DFT regression targets, SAGE-Net achieves the lowest MAE on eight targets when instantiated with different GNN backbones. These benchmarks cover bandgap, mechanical, dielectric, and transport-related properties, demonstrating strong or highly competitive performance against both structure-based and multimodal baselines. These results demonstrate that SGMP provides a transferable strategy for integrating crystallographic semantics into geometric message passing, establishing SAGE-Net as a general framework for deeply coupled multimodal materials learning.

## RESULTS AND DISCUSSION

## Description of model architecture

Figure 1a compares representative structure–description integration strategies. Unlike post-encoding fusion and latent-space alignment, SAGE-Net introduces crystallographic semantics during graph encoding through SGMP, thereby modulating intermediate node representations and subsequent message passing before graph pooling. As illustrated in Figure 1b, SAGE-Net comprises four components: a structural encoder, a text encoder, an SGMP module, and a gated late-fusion prediction head. The central design principle is to introduce crystallographic description-derived semantics into intermediate crystal-graph representations during structural encoding, rather than relying exclusively on shallow fusion after graph-level pooling.

For each material, the structural branch is implemented with a flexible GNN backbone, including CGCNN, ALIGNN, DenseGNN-Lite, and CartNet, to encode the geometric and chemical environments of crystalline materials. These backbones represent distinct geometric learning paradigms: CGCNN captures local crystal environments through crystal graph convolution; ALIGNN explicitly models bond-angle interactions using an additional line graph; DenseGNN-Lite enhances deep message passing through dense and residual information propagation to alleviate oversmoothing; and CartNet incorporates Cartesian geometric encoding to capture direction-dependent structural information in crystalline materials.[20,21,24,30] In parallel, the text branch utilizes MatSciBERT to encode standardized structure-derived crystallographic descriptions generated by RoboCrystallographer.[27,31] These descriptions provide complementary semantic cues (Table S1), including symmetry information, coordination environments, dimensionality, composition/site identity, and other chemically relevant priors.

To bridge atomic geometry with crystallographic descriptors, SAGE-Net adopts a dual-stage multimodal design comprising (i) node-level SGMP within the selected structural encoder and (ii) gated late fusion for adaptive modality balancing at the graph level. Through SGMP, crystallographic descriptors are converted into node-level gating signals that modulate atom-level updates during message passing. For example, symmetry cues, such as a space group or crystal-system label, can guide message passing toward symmetry-consistent local environments. Coordination descriptors such as "octahedral" or "tetrahedral" can emphasize features associated with corresponding polyhedral motifs. Dimensionality cues, such as "layered" or "two-dimensional," can strengthen interactions consistent with anisotropic connectivity. This design yields a unified materials representation that is both geometrically grounded and crystallographically informed. By combining semantically modulated atomistic features with global text representations, SAGE-Net captures interactions between local atomic environments and global crystallographic descriptors, providing a robust foundation for predicting bandgaps, mechanical responses, transport-related metrics, and crystal synthesizability.

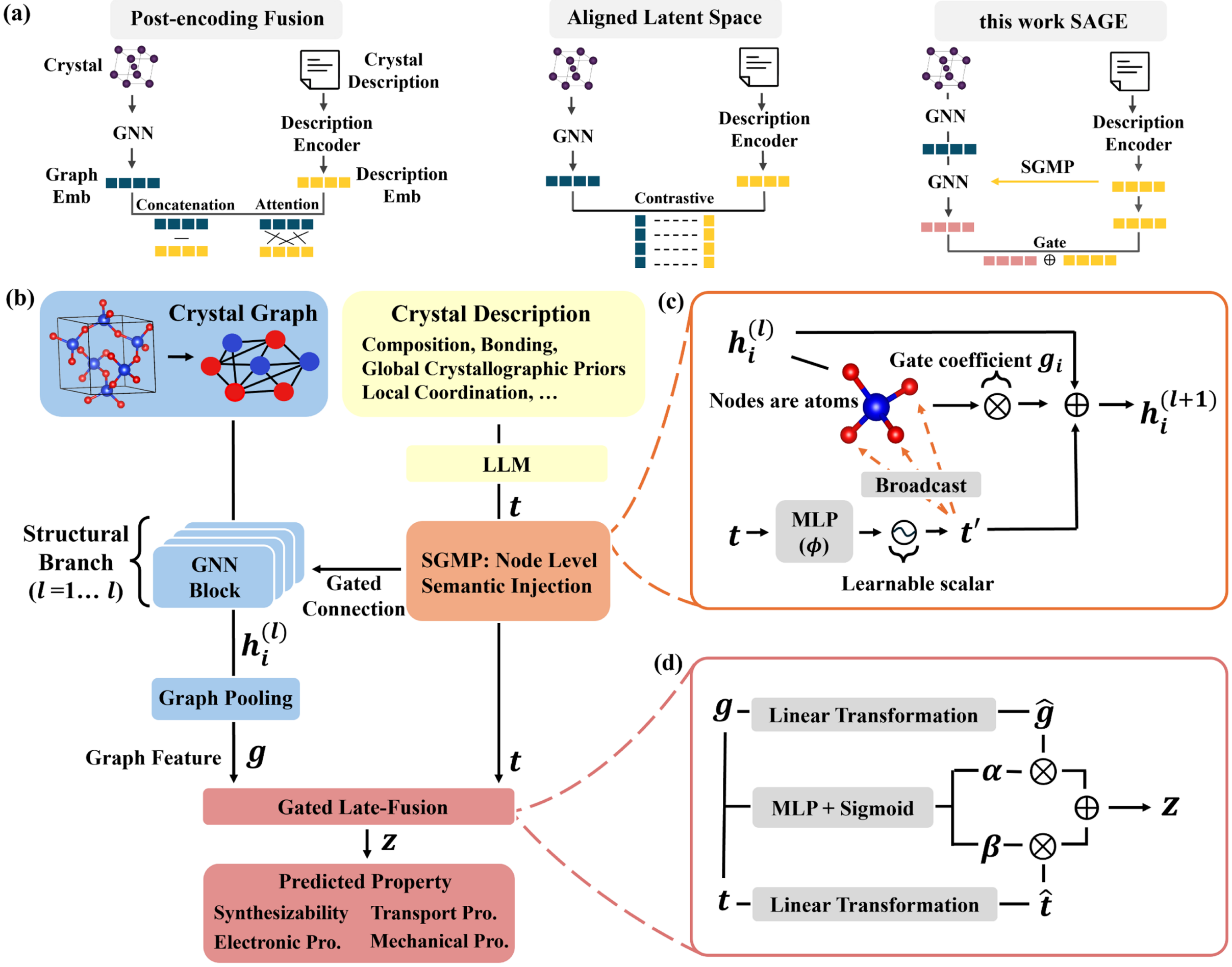

**Figure 1. The SAGE-Net architecture for semantics-augmented crystal property prediction. (a)** Comparison of representative structure–description integration strategies. **(b)** Overall workflow of the dual-stage multimodal framework. **(c)** Detailed mechanism of the SGMP module. **(d)** The gated late-fusion module.

### SGMP: node-level semantic injection

To achieve this conditional modulation (Figure 1c), let $h_i^{(l)} \in \mathbb{R}^{256}$ denote the feature of node $i$ after the $l$-th block of the selected GNN backbone and let $t \in \mathbb{R}^{64}$ be the pooled crystallographic-description embedding for the corresponding structure. We first project the text embedding to the node feature dimension via an MLP $\phi : \mathbb{R}^{64} \rightarrow \mathbb{R}^{256}$ and modulate its magnitude with a learnable scalar $\gamma$ initialized to 1.0:

$$t' = \gamma\phi(t) \quad (1)$$

The transformed text vector $t'$ is then broadcast to all nodes in the structure. A node-wise gate is computed from the concatenation of the node and broadcasted text features:

$$g_i = \sigma(W_g[h_i^{(l)} \,||\, t']) \in (0,1)^{256} \quad (2)$$

and node features are updated through a residual gated injection:

$$h_i^{(l+1)} = LN(h_i^{(l)} + g_i \odot t') \quad (3)$$

followed by dropout. Injecting semantics at an intermediate encoder depth allows subsequent message passing blocks to further refine text-modulated atom representations. Concretely, the text representation is projected into the node feature space and adaptively modulates each atomic update via a node-wise gate, allowing different atomic environments to respond selectively to the same global semantic cue. This conditional modulation enables structure-derived textual information to act as a crystallographic prior during representation formation rather than as a post hoc predictor. Additional architectural details (including gated late fusion, as well as optional fine-grained fusion and optional cross-modal attention), together with hyperparameters, are provided in the Methods.

### Gated late fusion

After semantic-guided encoding, we obtain a graph-level representation $g \in \mathbb{R}^d$ from the block of the selected GNN backbone and a text representation $t \in \mathbb{R}^d$ from the language encoder. We combine these modalities using a gated late-fusion module that adaptively balances their contributions (Figure 1d). Specifically, two lightweight gating networks map each modality to a scalar importance score:

$$\tilde{\alpha} = \sigma(f_g(g)),\ \tilde{\beta} = \sigma(f_t(t)) \tag{4}$$

where $\sigma(\cdot)$ denotes the sigmoid function and $f_g(\cdot)$ and $f_t(\cdot)$ are two-layer multilayer perceptrons. The weights are then normalized to enforce $\alpha + \beta = 1$:

$$\alpha = \frac{\tilde{\alpha}}{\tilde{\alpha}+\tilde{\beta}+\varepsilon},\ \beta = \frac{\tilde{\beta}}{\tilde{\alpha}+\tilde{\beta}+\varepsilon} \tag{5}$$

with $\varepsilon$ a small constant for numerical stability. Before fusion, both modalities are projected to a common latent space using linear transformations:

$$\hat{g} = W_g,\ \hat{t} = W_t \tag{6}$$

The fused representation is then computed as a convex combination:

$$z = \alpha\hat{g} + \beta\hat{t} \tag{7}$$

Finally, z is passed through a transformation block (linear layer, layer normalization, rectified linear unit (ReLU), and dropout) before the task-specific prediction head. This design allows the model to emphasize the modality that provides the most informative signal for a given crystal-description pair while keeping the fusion weights interpretable.

**Model comparison and analysis**

To evaluate the performance of SAGE-Net, we benchmarked it against seven representative baselines on ten regression targets from the JARVIS-DFT dataset, covering bandgaps, mechanical, dielectric, and transport-related properties. These baselines include four structure-based models, CGCNN, ALIGNN, DenseGNN-Lite, and CartNet, as well as three multimodal models, CrysMMNet, MultiMat, and Hybrid-LLM-GNN.[19-21,24,29,30] Details regarding the sources and sizes of the datasets are provided in the Methods section. As summarized in Table 1, SAGE-Net achieves the lowest MAE on eight out of ten regression targets when instantiated with different GNN backbones, demonstrating strong generality across diverse property classes. Specifically, SAGE-ALIGNN obtains the best performance for electric field gradient, SLME, and average hole mass; SAGE-CartNet achieves the lowest MAE for mBJ bandgap, OptB88vdW bandgap, and bulk modulus; and SAGE-DenseGNN-Lite performs the best for shear modulus and static dielectric constant. For the remaining two targets, Seebeck coefficient and spin–orbit spillage, CrysMMNet and Hybrid-LLM-GNN give the lowest MAEs, respectively, while SAGE-Net remains competitive. Parity plots for all ten regression targets are provided in Figures S1 and S2.

**Table 1. Comparison of prediction performance for SAGE-Net and baseline models across ten JARVIS-DFT regression targets.** Mean absolute error (MAE) is reported for each property, with lower values indicating better performance. The asterisk (*) denotes the best-performing model for each target. Benchmark results in this table were obtained using the official dataset splits.

| Mean Absolute Error (MAE) | | | | | |
|---|---|---|---|---|---|
| Model | Electric field gradient (V/Å$^2$) | Bandgap (mBJ) (eV) | SLME (%) | Seebeck (μV/K) | Average hole mass ($m_0$) |
| CGCNN | 24.670 | 0.413 | 5.660 | 49.317 | 0.157 |
| ALIGNN | 19.121 | 0.310 | 4.521 | 40.921 | 0.124 |
| DenseGNN-Lite | 21.202 | 0.267 | 4.793 | 44.204 | 0.128 |
| CartNet | 21.708 | 0.251 | 4.495 | 41.712 | 0.125 |
| CrysMMNet | 19.021 | 0.271 | 4.500 | 40.120* | 0.125 |
| MultiMat | 19.714 | 0.291 | 4.670 | 40.990 | 0.132 |
| Hybrid-LLM-GNN | 21.729 | 0.313 | 4.801 | 45.725 | 0.118 |
| SAGE-CGCNN | 22.404 | 0.302 | 4.853 | 48.768 | 0.132 |
| SAGE-ALIGNN | 18.472* | 0.257 | 4.281* | 42.438 | 0.114* |
| SAGE-DenseGNN-Lite | 19.246 | 0.243 | 4.498 | 41.809 | 0.120 |
| SAGE-CartNet | 20.811 | 0.234* | 4.396 | 40.552 | 0.121 |

| Mean Absolute Error (MAE) | | | | | |
|---|---|---|---|---|---|
| Model | Static dielectric constant (/) | Spillage (/) | Bandgap (OptB88vdW) (eV) | Bulk moduli (GPa) | Shear moduli (GPa) |
| CGCNN | 36.054 | 0.397 | 0.200 | 14.120 | 11.980 |
| ALIGNN | 23.731 | 0.351 | 0.142 | 10.400 | 9.481 |
| DenseGNN-Lite | 23.368 | 0.361 | 0.140 | 9.511 | 8.900 |
| CartNet | 24.307 | 0.373 | 0.117 | 8.776 | 8.826 |
| CrysMMNet | 24.015 | 0.348 | 0.124 | 10.308 | 9.587 |
| MultiMat | 24.187 | 0.363 | 0.155 | 10.669 | 9.605 |
| Hybrid-LLM-GNN | 24.532 | 0.343* | 0.144 | 17.979 | 11.498 |
| SAGE-CGCNN | 25.357 | 0.355 | 0.151 | 12.235 | 10.293 |
| SAGE-ALIGNN | 23.057 | 0.348 | 0.122 | 10.420 | 9.305 |
| SAGE-DenseGNN-Lite | 22.908* | 0.349 | 0.125 | 9.232 | 8.772* |
| SAGE-CartNet | 23.595 | 0.360 | 0.106* | 8.508* | 8.970 |

To provide a more global view of model performance, Figure 2 summarizes the task-wise rankings and relative improvements across the ten JARVIS-DFT regression targets. Figure 2a shows that SAGE-CartNet and SAGE-ALIGNN each rank first on three targets and second on two additional targets, while SAGE-DenseGNN-Lite ranks first on two targets and second on one target. Together with the results in Table 1, these rankings indicate that no single geometric backbone dominates all property targets, either before or after SGMP integration. This reflects the different inductive biases of the backbones, including angle-aware message passing in ALIGNN, Cartesian geometric encoding in CartNet, and dense information propagation in DenseGNN-Lite. Consistently, the average-rank analysis in Figure 2b identifies SAGE-CartNet as the top-ranked model with an average rank of 3.20, followed by SAGE-DenseGNN-Lite and SAGE-ALIGNN with average ranks of 3.30 and 3.35, respectively. These results support SAGE-Net as a flexible SGMP-based framework that injects crystallographic semantics into different geometric encoders while preserving their target-dependent strengths.

Figure 2c further compares each SAGE-Net model with its corresponding structure-only backbone. The SGMP-enhanced models generally improve over their base models across most targets, confirming that crystallographic description-derived information provides useful priors beyond the original crystal-graph representation. The improvement is particularly evident for SAGE-CGCNN, which shows broad gains relative to CGCNN, and for SAGE-CartNet and SAGE-ALIGNN, which achieve high overall rankings while improving several properties, viz. bandgaps, dielectric, mechanical, and transport-related properties. Although the magnitude of improvement varies across properties and backbones, these backbone-level comparisons support the view that SGMP is not tied to a specific GNN architecture.

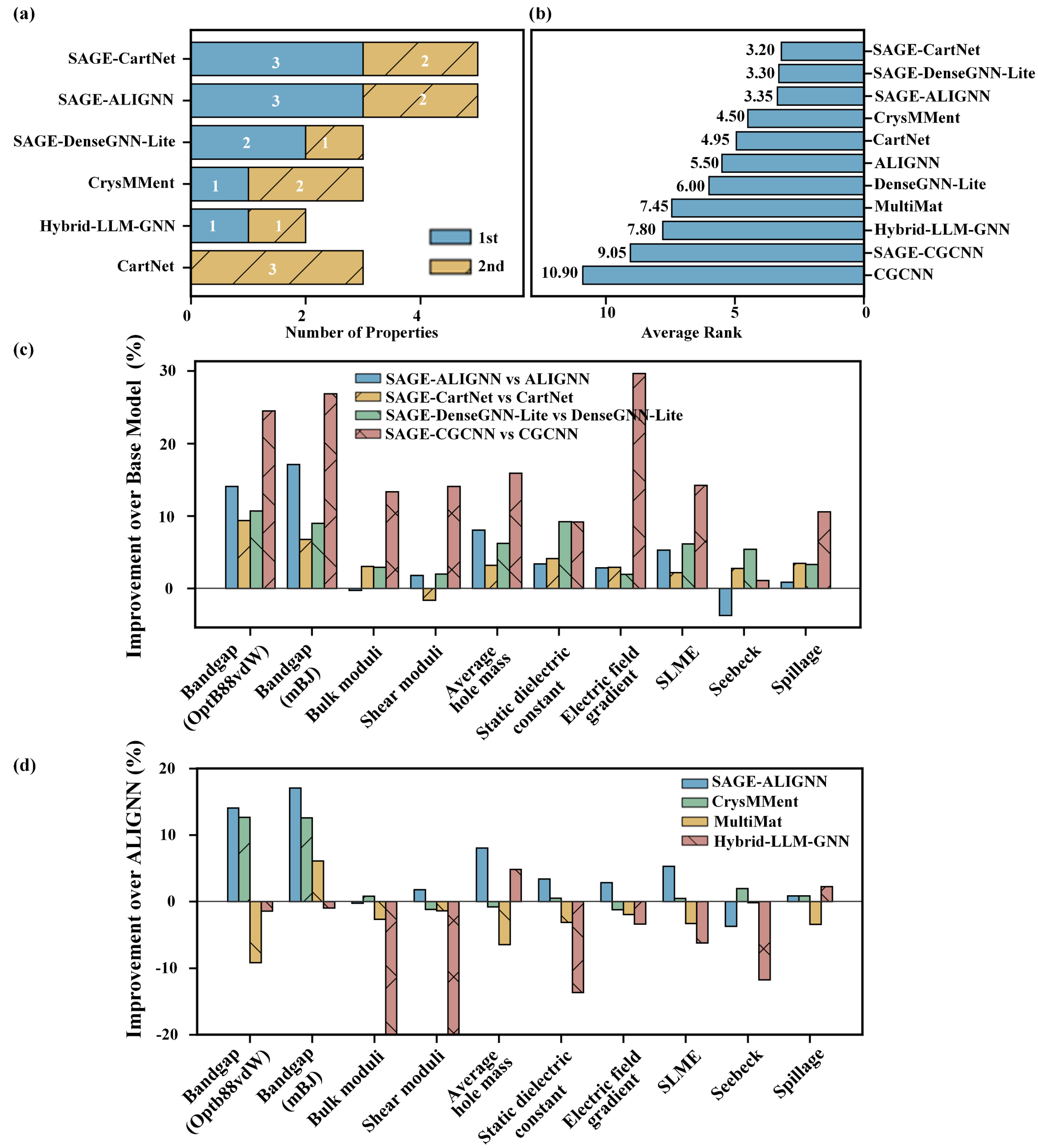

**Figure 2. Overall benchmark summary of SAGE-Net and baseline models across ten JARVIS-DFT regression targets. (a)** Number of properties for which each model ranks first or second in terms of test MAE. Solid bars denote first-place counts, and hatched bars denote second-place counts. **(b)** Average rank of each model across the ten regression targets, where a lower rank indicates better overall performance.

**(c)** Percentage improvement of each SAGE-Net instantiation over its corresponding structure-only backbone, including SAGE-CGCNN versus CGCNN, SAGE-ALIGNN versus ALIGNN, SAGE-DenseGNN-Lite versus DenseGNN-Lite, and SAGE-CartNet versus CartNet. **(d)** Percentage improvement over the ALIGNN baseline for SAGE-ALIGNN and representative multimodal baselines, including CrysMMNet, MultiMat, and Hybrid-LLM-GNN. Positive values indicate lower MAE than the reference baseline, whereas negative values indicate performance degradation.

Moreover, Figure 2d compares SAGE-ALIGNN with representative multimodal baselines using ALIGNN as the reference model. SAGE-ALIGNN shows more stable positive improvements over ALIGNN across most targets, whereas MultiMat and Hybrid-LLM-GNN exhibit performance degradation on certain properties. CrysMMNet performs well on some targets, but its improvements are less uniform across the benchmark suite. These observations indicate that simply introducing textual information through post-encoding fusion, latent-space alignment, or representation-level multimodal interaction does not necessarily lead to consistent gains. However, SAGE-Net benefits from injecting crystallographic semantic information into atom-level message passing, allowing structure-derived descriptions to shape crystal-graph representations before graph-level pooling.

Because RoboCrystallographer descriptions are generated from crystal structures, we performed two tests to verify that the improvement arises from structure-consistent semantics rather than incidental effects of adding a text branch. In a structure–text mismatch test, we randomly permuted the descriptions across the test set while keeping each crystal graph fixed. For SAGE-ALIGNN, this perturbation increases the mBJ bandgap MAE from 0.246 eV with matched text to 0.311 eV with mismatched text (Figures S3a and S3b), nearly reverting to the structure-only ALIGNN baseline (0.310 eV). This indicates that unaligned textual information does not provide useful gains and may weaken semantic-guided message passing. We further replaced the MatSciBERT representation with an explicit crystallographic prior vector computed from CIF, including symmetry, dimensionality, and coordination-number statistics, and injected it using the same gating mechanism. This explicit-prior baseline partially recovers the improvement, reaching an mBJ MAE of 0.279 eV (Table S2), but remains inferior to the matched-text model. These trials show that the performance gain is not merely caused by additional input features or increased model capacity, but depends on aligned semantic information coupled to atom-level message passing.

**Ablation study**

To quantify the contribution of each multimodal component, we performed an ablation study on four representative property prediction tasks (Table 2). The baseline is a gated late-fusion model, in which graph and text representations are independently encoded and adaptively combined only at the prediction stage. Model 1 adds Semantic-Guided Message Passing (SGMP), where crystallographic description-derived semantics modulate atom-level updates through semantic gating during geometric encoding. As shown in Table 2, enabling SGMP yields the most consistent improvements. We further evaluate optional modules both individually and in combination with SGMP. Model 2 adds fine-grained fusion only, and Model 3 adds cross-modal attention only. For SLME, the best result is obtained when SGMP is combined with fine-grained fusion (Model 1+2), reducing the MAE from 4.332% to 4.229%, while Model 1 alone still provides a modest improvement (4.276%). By contrast, enabling fine-grained fusion or cross-modal attention without SGMP (Model 2 or Model 3) does not yield systematic benefits, and can even degrade performance. An end-to-end schematic of these module configurations is provided in Figure S4. Overall, the ablation studies indicate that textual semantics are the most effective when introduced as chemically meaningful constraints during message passing, whereas additional fusion/attention mechanisms provide target-dependent and generally incremental improvements.

**Table 2. Ablation study of model components on four property prediction tasks.** The baseline corresponds to a gated late-fusion model, in which graph and text representations are independently encoded and adaptively combined at the prediction stage. Model 1 adds SGMP only. Model 2 adds fine-grained fusion only. Model 3 adds cross-modal attention only. An asterisk (*) indicates the best performance.

| | Mean Absolute Error (MAE) | | | |
|---|---|---|---|---|
| Model | Spillage (/) | Bandgap (mBJ) (eV) | SLME (%) | Average hole mass ($m_0$) |
| Baseline | 0.347 | 0.288 | 4.332 | 0.121 |
| +Model (1) | 0.346* | 0.246* | 4.276 | 0.119* |
| +Model (1+2) | 0.349 | 0.258 | 4.229* | 0.126 |
| +Model (1+2+3) | 0.353 | 0.251 | 4.314 | 0.119* |
| +Model (2) | 0.353 | 0.282 | 4.421 | 0.119* |
| +Model (2+3) | 0.347 | 0.269 | 4.308 | 0.122 |

**Mechanistic insights into semantic conditioning**

To elucidate the origin of the observed performance gains, we analyze how crystallographic descriptions influence the evolution of crystal-graph representations during message passing. Unless otherwise mentioned, this section uses SAGE-ALIGNN, trained on the mBJ bandgap task as a representative example. To examine when structure-derived textual descriptors begin to affect the crystal representation, we tracked the layer-wise consistency between atom-level crystal-graph features and the corresponding crystallographic description features (Figure 3a). In the early ALIGNN blocks (Blocks 1–3), the consistency remains low, indicating that the model first focuses on establishing local bonding geometry, coordination environments, and short-range structural connectivity. After SGMP is introduced at Block 3, the consistency increases sharply, suggesting that crystallographic descriptors such as symmetry, coordination motifs, dimensionality, and site information begin to guide the evolution of atom-level representations. In deeper ALIGNN blocks, the consistency gradually relaxes toward Block 8, implying that the injected crystal-level semantic information is not simply copied into the graph representation, but is further reorganized under local geometric and bonding constraints. This trend agrees with the injection-depth ablation in Table S3, where intermediate SGMP insertion gives the lowest prediction error, whereas too-early or repeated semantic injection leads to inferior performance.

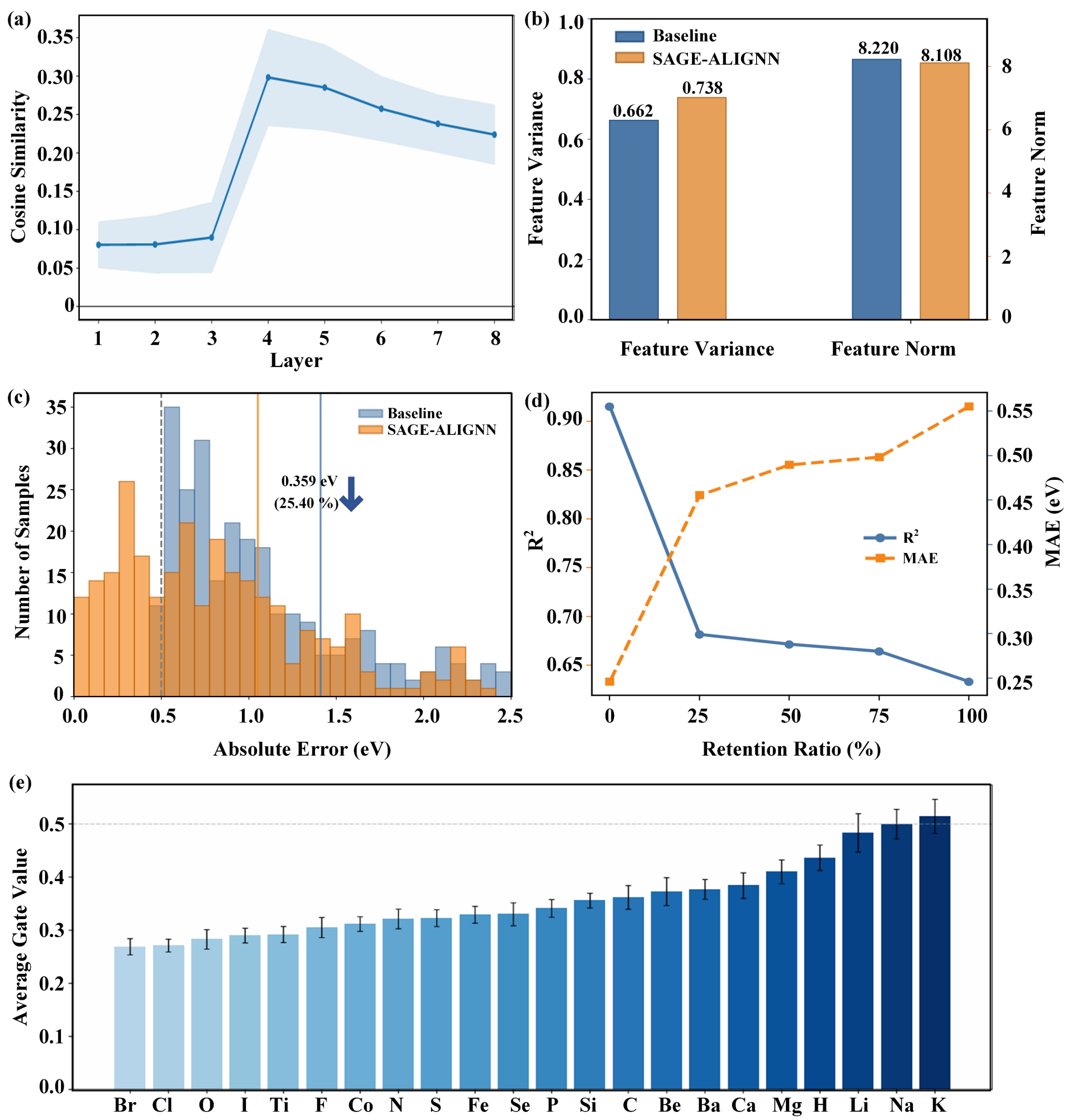

**Figure 3. Crystallographic descriptions modulate crystal-graph representations to resolve hard samples and enhance physical interpretability. (a)** Layer-wise cosine similarity (node features vs text embeddings). **(b)** Feature statistics, **(c)** Hard-sample error distribution/MAE shift. **(d)** Test-time keyword retention/masking (MAE and $R^2$). **(e)** Element-wise average gating coefficients α.

Taken together, these results suggest a consistent design rule: crystallographic descriptions are the most beneficial when introduced after the crystal graph has already established basic local structural information. In the early ALIGNN blocks, the model primarily learns local coordination geometry, bond connectivity, and short-range atomic environments from the crystal graph. Introducing global crystallographic descriptors too early may disturb this initial construction of local structural representations. In contrast, mid-level SGMP allows symmetry, coordination, dimensionality, and site-related descriptors to act as crystallographic priors, which can then be further refined by subsequent message-passing layers under local bonding and geometric constraints. This trajectory indicates that the information carried by the crystal descriptions is not simply copied into the graph representation, but is progressively integrated with local structural environments during message passing. Consistent with this behavior, SAGE-ALIGNN modifies the distribution of learned crystal representations (Figure 3b), increasing the feature variance from 0.662 to 0.738 while maintaining a comparable feature norm, suggesting improved distinction among different chemical and coordination environments.

The practical impact of this structure–description coupling is most apparent for challenging crystals (Figure 3c). For "hard" samples where the CrysMMNet baseline gives large errors (> 0.500 eV, n = 287), SAGE-ALIGNN shifts the error distribution toward smaller values and reduces the subset MAE by 0.359 eV, corresponding to a 25.4% reduction. To examine how much the model depends on the crystallographic description, we performed a test-time keyword retention analysis (Figure 3d). Because the text input is constructed from curated chemistry and crystallography keywords, the retention ratio directly controls the amount of available structural semantic information at inference. The prediction accuracy recovers monotonically as more keywords are retained, with $R^2$ increasing from 0.630 to 0.915. This confirms that the improvement is strongly linked to the availability of structure-consistent crystallographic descriptors.

Finally, the element-resolved gating coefficients show systematic chemical trends across different atomic species (Figure 3e). Alkali metals and hydrogen exhibit higher gate values ($\alpha > 0.450$), indicating stronger modulation by crystallographic descriptions, whereas halogens and more strongly coordinated species show lower gate values ($\alpha < 0.350$). These differences suggest that SAGE-ALIGNN does not apply semantic information uniformly to all atoms. Instead, the degree of description-guided modulation depends on the chemical identity and local coordination environment, consistent with the chemically heterogeneous nature of crystalline materials.

**Semantic selectivity and interpretability**

While the preceding analyses show that crystallographic descriptions enter and reshape crystal-graph representations, it remains important to determine which types of descriptors are actually leveraged by the model. As shown in Figure 4a, global crystallographic priors (crys.) and structure descriptors (stru.) are strongly enriched in attention relative to their corpus prevalence. This indicates that SAGE-ALIGNN preferentially amplifies low-frequency but diagnostically important crystallographic cues beyond what would be expected from token frequency alone. The attention distributions in Figure 4b further support this observation: crys. tokens exhibit relatively high attention weights with concentrated distributions, consistent with the stable use of global symmetry and lattice-level information across samples.

Because attention enrichment reflects selective use rather than marginal predictive importance, we next quantified functional sensitivity via group-wise masking on the full test set (Figure 4c). For each category, tokens were masked when present while all other inputs were kept unchanged; we report both the resulting increase in MAE and the category coverage. Masking composition and site identity (comp.) produces the largest degradation in bandgap MAE ($\Delta$MAE = 28.1%), followed by local coordination and polyhedral environment (coor.) and global crystallographic priors (crys.) ($\Delta$MAE = 15.6%). In contrast, shape and geometric appearance (shap.) has negligible impact and serves as an internal negative control. Consistent with these results, the association between attention enrichment and masking impact is weak and not statistically significant across categories (Spearman $\rho = 0.357$, $p = 0.432$; Figure 4d), underscoring that attention selectivity and performance sensitivity capture complementary, non-interchangeable aspects of semantic conditioning. Additionally, we further probed semantic conditioning by counterfactual swapping of structure-type tokens while keeping the crystal structures fixed; the induced prediction shifts are reversible across multiple prototype pairs (Figure S5). This intervention is intended as a sensitivity probe and does not imply a physically realizable prototype transformation.

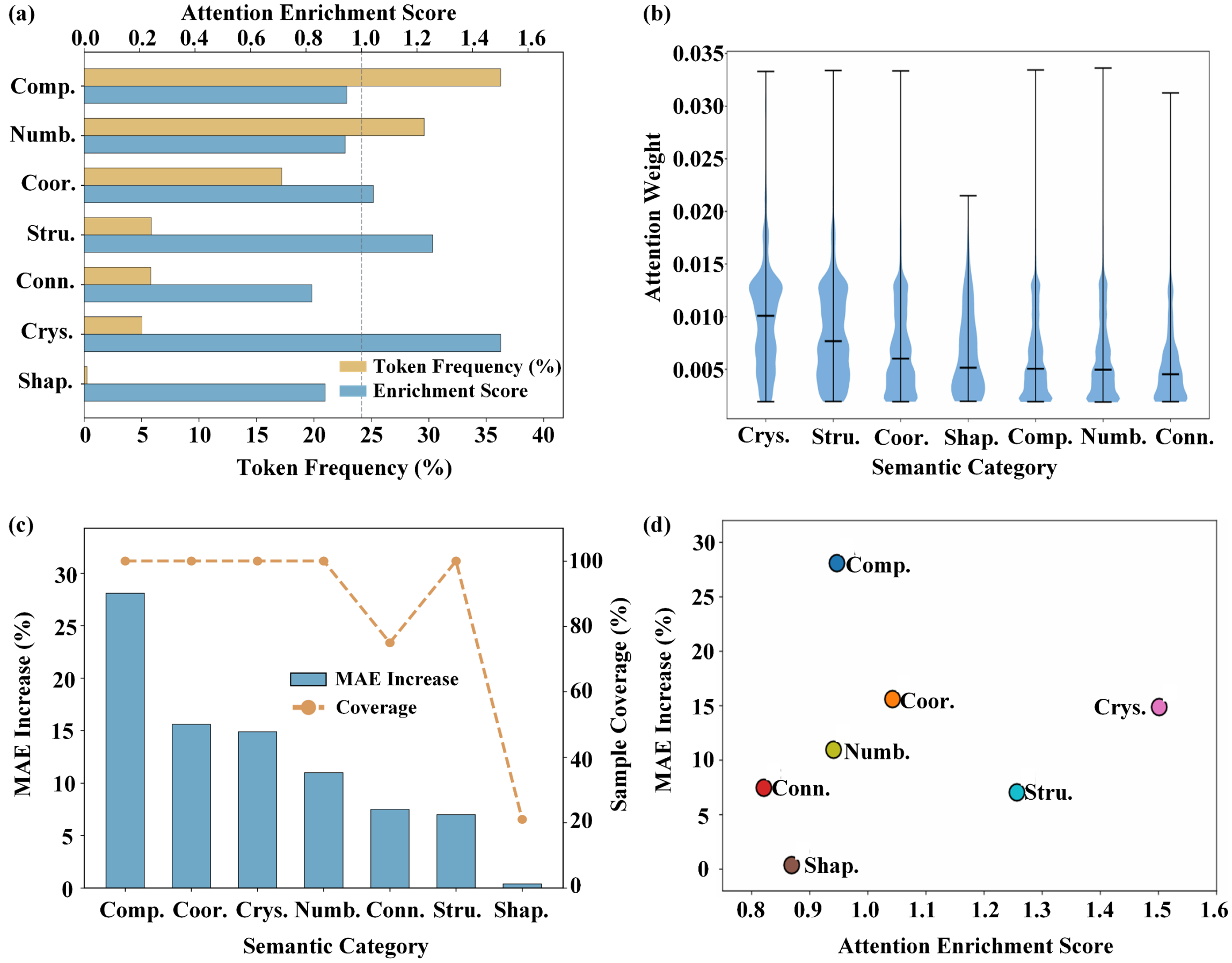

**Figure 4. Category-wise ablation for the text branch. (a)** Attention enrichment score (light) versus corpus frequency (dark) for seven semantic super-categories. **(b)** Distributions of token-level attention weights across categories. **(c)** Group-wise masking analysis reporting the relative MAE increase (bars) and category coverage (dashed line). **(d)** Relationship between attention enrichment and masking impact across categories.

The practical value of deeply coupling text semantics with geometric learning is highlighted by the mBJ bandgap prediction for $Rb_2AuGaI_6$ (Figure S6a). This preference for crystallographic priors, highlighted by the enrichment and attention analyses in Figure 4a–d, is further illustrated by this representative case study (Figure S6a, b). For this quaternary double perovskite, the multimodal late-fusion baseline, which combines text and structure only at the final prediction stage, still incurs a substantial error (0.269 eV).

This behavior is consistent with the chemical complexity of double perovskites, where cation ordering and heterogeneous coordination environments can yield similar local geometries but distinct electronic responses. In contrast, SAGE-ALIGNN reduces the error to 0.002 eV by leveraging explicit crystallographic descriptors such as “Fm-3m” together with coordination and polyhedral motifs such as “cuboctahedra,” which encode chemically meaningful constraints on site symmetry and local environments. Importantly, the atom–token attention heatmap (Figure S6b) provides chemically interpretable grounding between language and atomic environments. Tokens associated with polyhedral language align with the site chemistry expected for a double perovskite, attention linked to the B-site cations (Au and Ga) concentrates on coordination descriptors such as “octahedra,” consistent with $BX_6$ type units, whereas Rb aligns with A-site polyhedral motifs such as “cuboctahedra,” which describe the larger coordination cage characteristic of alkali A-site cations. The attention also highlights how different semantic cues play complementary roles, with space-group information providing a global symmetry constraint and polyhedral descriptors specifying local coordination geometry. Together, these results indicate that SAGE-ALIGNN learns interpretable cross-modal correspondences that connect crystallographic language to specific atomic environments, thereby improving predictions for structurally complex materials beyond what late fusion (concatenation) alone captures. Additional interpretability case studies for $Na_4BeAlSi_4ClO_{12}$ (six elements compound) are provided in Figure S6c, d.

**Application to synthesizability screening**

Having established that SGMP improves property prediction by coupling crystallographic descriptions with crystal-graph learning, we next examine whether this mechanism also benefits downstream materials screening. To demonstrate the utility of SAGE-Net in practical materials discovery workflows, we evaluated it on a binary synthesizability screening task using the dataset described in the Methods section. The test set spans a broad chemical and structural space, is dominated by ternary and quaternary compounds, and covers all seven crystal systems (Figure 5a). The confusion matrix in Figure 5b indicates balanced performance, correctly identifying 3970 unsynthesizable and 4084 synthesizable candidates. For screening, the relatively low false-negative count (120) compared with false positives (208) is particularly valuable, because missing experimentally accessible candidates is often more costly than advancing a manageable number of false positives to subsequent validation.

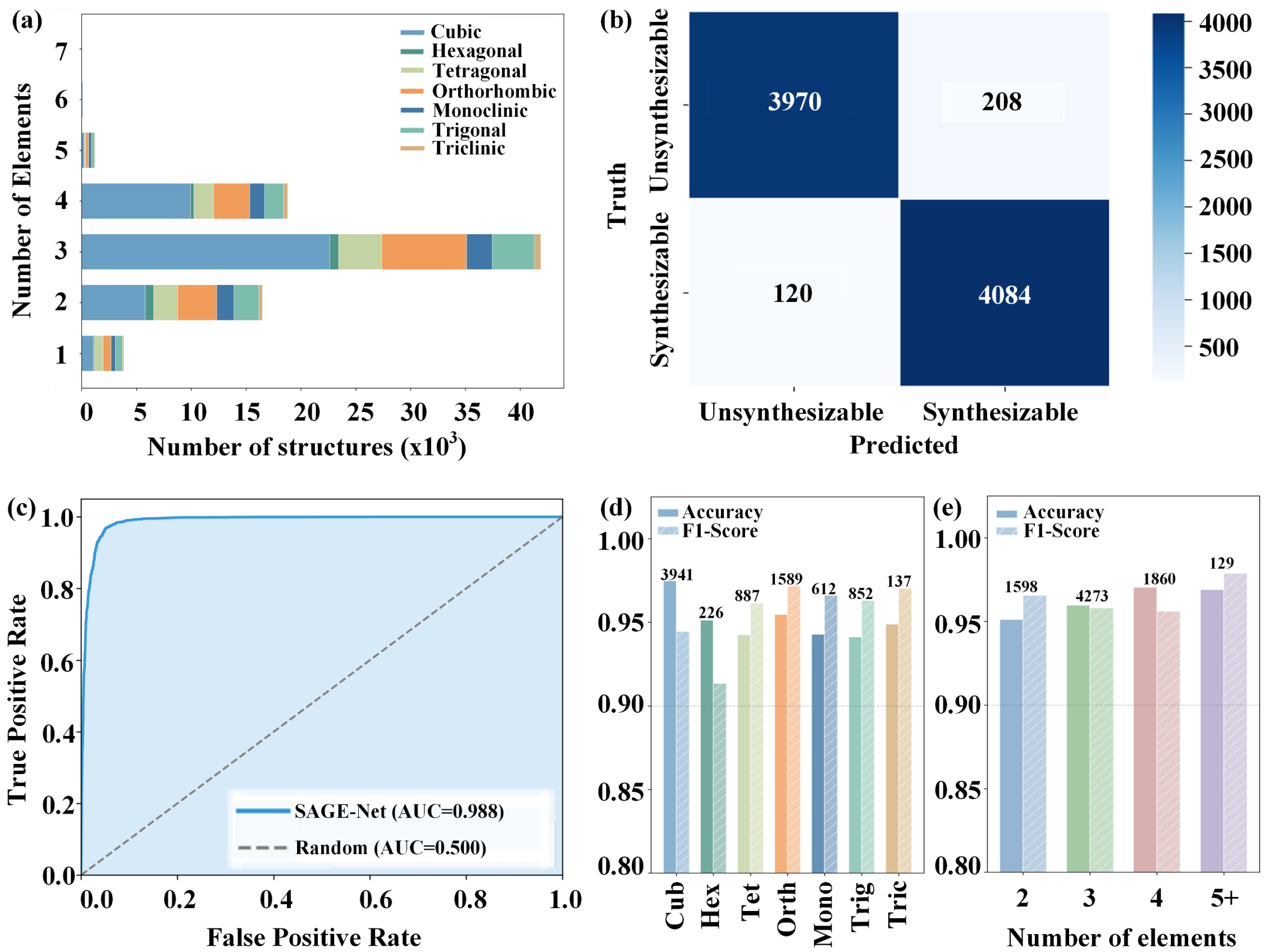

**Figure 5. Synthesizability screening performance of SAGE-Net. (a)** Dataset distribution by number of elements and crystal system. **(b)** Confusion matrix for binary synthesizability classification. **(c)** ROC curve with AUC = 0.988. **(d)** Accuracy and F1-score stratified by crystal system (Cub, cubic; Hex, hexagonal; Tet, tetragonal; Orth, orthorhombic; Mono, monoclinic; Trig, trigonal; Tric, triclinic). **(e)** Accuracy and F1-score stratified by number of constituent elements.

This diversity provides a stringent assessment of generalization beyond simple compositions or high-symmetry subsets. SAGE-Net shows strong global discrimination, achieving an ROC–AUC of 0.988 (Figure 5c). We further examined robustness with respect to structural and compositional complexity. Across crystal systems from cubic to triclinic, SAGE-Net maintains consistently high accuracy (> 94.0%) and F1-Score (Figure 5d), indicating that performance is not confined to a narrow symmetry class. Likewise, accuracy and F1 remain stable as the number of constituent elements increases from binary to quinary (5+) compositions (Figure 5e), suggesting resilience to the combinatorial complexity of multicomponent chemistry.

Finally, we benchmarked SAGE-ALIGNN against its geometric backbone ALIGNN and the language-model-based method CSLLM (Table 3).[20,32] Compared with ALIGNN, SAGE-ALIGNN improves the accuracy from 94.80% to 96.09% and increases the recall from 96.98% to 97.15%. These gains support the premise that text-derived chemical descriptors provide information that complements structure alone. Relative to CSLLM, SAGE-Net achieves comparable overall performance while delivering the highest recall among the evaluated methods (97.15%). Taken together, these results indicate that SAGE-Net is well suited for high-throughput screening, where prioritizing recall helps minimize missed synthesizable candidates.

**Table 3. comparison with ALIGNN (graph baseline and SAGE-ALIGNN backbone) and CSLLM.**

| Model | Accuracy | Recall | F1-Score | Precision |
|---|---|---|---|---|
| ALIGNN | 94.80 | 96.98 | 94.92 | 92.96 |
| CSLLM | 96.22 | 96.87 | 96.24 | 95.62 |
| SAGE-ALIGNN | 96.09 | 97.15 | 96.14 | 95.15 |

## CONCLUSIONS

In summary, we developed SAGE-Net, a flexible multimodal framework that couples structure-derived crystallographic descriptions with crystal-graph message passing for crystalline materials. SAGE-Net introduces crystallographic semantic information during atom-level updates, allowing descriptors such as symmetry, dimensionality, space group, site identity, and coordination motifs to shape local structural representations during encoding. By integrating SGMP with multiple GNN backbones, including CGCNN, ALIGNN, DenseGNN-Lite and CartNet, SAGE-Net provides a transferable strategy for enhancing geometric representation learning.

Across benchmarks spanning bandgaps, mechanical, dielectric, and transport-related properties, SAGE-Net improves its corresponding structure-only backbones on most targets when instantiated with different geometric encoders, and achieves the lowest MAE on eight out of ten JARVIS-DFT regression targets. The gains are particularly evident for challenging cases where structure-only representations or post-encoding multimodal fusion provide insufficient structure–property discrimination. Beyond predictive accuracy, SAGE-Net offers chemically interpretable cross-modal grounding. Attention enrichment, masking, and case-study analyses show that the model preferentially leverages crystallographic priors rather than merely frequent tokens, and learns intuitive correspondences between polyhedral or symmetry descriptors and the associated atomic environments. SAGE-Net also demonstrates practical utility for materials discovery, achieving strong performance in synthesizability screening with ROC–AUC = 0.988 and 97.15% recall, supporting conservative down-selection where minimizing missed synthesizable candidates is a primary objective.

More broadly, this work establishes a general strategy for multimodal learning in materials science by treating crystallographic language not as auxiliary metadata, but as an actionable source of chemical and structural priors that can be grounded in crystal graphs. By extending this principle to richer materials knowledge sources, such as synthesis narratives, processing conditions, experimental metadata, and multimodal pretraining, future models may move beyond static structure–property prediction toward more context-aware, experimentally grounded, and transferable materials discovery.

## METHODS

We evaluated SAGE-Net on ten regression targets from the JARVIS-DFT dataset, whose reference properties were computed using density functional theory (DFT) in Vienna Ab initio Simulation Package (VASP).[33] Most targets were obtained with the optB88vdW functional,[34] while mBJ calculations were used to provide improved evaluation of bandgap where available.[35] Dielectric-related labels were taken as the static dielectric constant computed via Density Functional Perturbation Theory (DFPT), including both electronic and ionic contributions.[36] Thermoelectric quantities were calculated using BoltzTrap software under the constant relaxation-time approximation,[37] and spin–orbit spillage was computed from the difference between wavefunctions with and without spin–orbit coupling.[38] The mBJ frequency-dependent dielectric function is used to calculate the maximum efficiency limited by the spectrum.[39] For benchmark comparisons, we used the official train/validation/test splits of the corresponding JARVIS-DFT regression targets to ensure comparability with previously reported baselines. The validation set was used for early stopping and hyperparameter selection, and all benchmark results were evaluated on the held-out test set. The exact numbers of samples in each official split are provided in Table S4. Unless otherwise specified, all other experiments, including ablation studies, control experiments, interpretability analyses, and supplementary evaluations, used train/validation/test splits randomly generated with seed 42 where applicable.

Synthesizable (positive) structures were obtained from the Materials Project (MP) by selecting entries annotated with Inorganic Crystal Structure Database (ICSD) tags (experimentally reported structures).[40-42] Non-synthesizable (negative) structures were adopted from the dataset released in Crystal Synthesis Large Language Models (CSLLM),[32] where negatives are defined as low-likelihood crystals selected using a PU-learning crystallization-likelihood score (CLscore).[33,41,43-46] To standardize structural complexity, we retained ordered structures with $\leqslant 40$ atoms and $\leqslant 7$ elements using strict filters and performed stratified sampling by crystal system. We note that the negative labels are proxy labels derived from CLscore and may contain noise; therefore, we emphasize recall-oriented operating points for conservative screening.[32]

Since SAGE-Net leverages structure-derived textual descriptors as an additional modality, we first describe how these descriptions are generated and standardized. For each crystal structure, we generated a natural-language crystallographic description using RoboCrystallographer.[27] We then applied two preprocessing steps to reduce linguistic redundancy and redundant local geometric details: (i) removal of stopwords and function words using a predefined stopword lexicon,[47] and (ii) removal of local-detail sentences that explicitly report short-range geometric measurements, such as specific bond lengths and bond angles, typically accompanied by Å or ° units.[48] As a result, the average number of tokens per description was reduced by 57.4% (Figure S7). This preprocessing was designed to encourage the text branch to focus on higher-level crystallographic descriptors, such as symmetry, dimensionality, coordination motifs, and site-related information, while leaving explicit local geometric measurements to be captured primarily by the crystal-graph encoder. Following prior observations that filtering connective words and overly local geometric statements can improve predictive performance,[47,48] we evaluated the effect of this preprocessing using SAGE-DenseGNN-Lite across ten JARVIS-DFT property prediction tasks. The preprocessed descriptions yield lower MAE on seven out of ten targets, supporting their use in the main experiments; detailed ablation results are provided in Table S5. Overall, the preprocessed descriptions provide a more compact and generally robust text modality for SAGE-CartNet and are therefore used in the main experiments.

Fine-grained fusion for atom–token interpretability. In addition to SGMP, we implemented an optional fine-grained fusion module to resolve token-level interactions with atomic environments and to enable atom–token attention analysis. Let $h_i \in \mathbb{R}^{d_n}$ denote the representation of atom $i$ and $u_j \in \mathbb{R}^{d_t}$ denote the MatSciBERT token embedding for token $j$. We first project atomic and token features into a shared latent space of dimension $d_h$:

$$\tilde{h}_i = W_n h_i,\ \tilde{u}_j = W_t u_j \quad (8)$$

We then compute atom-to-token multi-head attention, where atoms serve as queries and tokens serve as keys and values:

$$Q = \tilde{H} W_Q,\ K = \tilde{U} W_K,\ V = \tilde{U} W_V,\ A = softmax(\frac{QK^{\top}}{\sqrt{d_h}}),\ C = AV \quad (9)$$

Padding tokens are masked before the softmax operation to ensure that attention weights are normalized only over valid tokens. The resulting token-conditioned context $C$ is projected back to the node-feature dimension to yield a text context vector for each atom. Finally, we integrate this context through a gated residual update:

$$\alpha_i = \sigma(W_g[h_i \| c_i]),\ h_i' = LayerNorm(h_i + \alpha_i \odot c_i) \quad (10)$$

where $\sigma(\cdot)$ is the sigmoid function, $\|$ denotes concatenation, and $\odot$ denotes elementwise multiplication. This design allows each atom to attend to different tokens and selectively incorporate token-level chemical information, providing a mechanistic basis for atom–token interpretability while keeping the module optional.

Cross-modal attention. We implemented an optional cross-modal attention block to enable bidirectional interaction between global graph and text representations. Let $g \in \mathbb{R}^{d_g}$ denote the pooled graph embedding from the geometric encoder and $t \in \mathbb{R}^{d_t}$ denote the global text embedding. The module performs two symmetric operations: graph-to-text (g2t) and text-to-graph (t2g), each implemented with multi-head scaled dot-product attention in a shared hidden space of dimension $d_h$.

For G2T, the graph embedding serves as the query and the text embedding serves as the key and value:

$$Q_{g2t} = W_{g2t}^{Q} g,\ K_{g2t} = W_{g2t}^{K} t,\ V_{g2t} = W_{g2t}^{V} t,$$

$$c_{g2t} = MHAttn\left(Q_{g2t}, K_{g2t}, V_{g2t}\right) = softmax\left(\frac{Q_{g2t} K_{g2t}^{\top}}{\sqrt{d_h}}\right) V_{g2t} \quad (11)$$

The resulting context is projected back to the graph dimension and added residually:

$$g' = LayerNorm(g + W_g^{O} c_{g2t}) \quad (12)$$

For T2G, the roles are reversed, with text as the query and graph as key and value:

$$Q_{t2g} = W_{t2g}^{Q} t,\ K_{t2g} = W_{t2g}^{K} g,\ V_{t2g} = W_{t2g}^{V} g,$$

$$c_{t2g} = MHAttn\left(Q_{t2g}, K_{t2g}, V_{t2g}\right),\ t' = LayerNorm(t + W_t^{O} c_{t2g}) \quad (13)$$

Here MHAttn denotes multi-head attention with $H$ heads, where each head applies scaled dot-product attention and the head outputs are concatenated and linearly projected. The enhanced embeddings $(g', t')$ are then passed to the downstream fusion and prediction blocks. This design allows the model to adjust the global structural representation using text context and, conversely, refine the text representation using global structural information.

For SAGE-ALIGNN, we optimized the mean squared error (MSE) loss with the AdamW optimizer (learning rate $5 \times 10^{-4}$, weight decay $1 \times 10^{-3}$) using a batch size of 64. Training was run for up to 100 epochs with early stopping based on validation performance (patience = 30 epochs). For SAGE-Net, SGMP was applied after ALIGNN layer 3. We use 0-based indexing in code; thus, middle_fusion_layers=2 corresponds to inserting SGMP after the third ALIGNN block. SAGE-Net was implemented in PyTorch.[49] Crystal structures were represented as graphs using GDL,[50] with structure parsing and neighbor construction handled by the JARVIS toolkit.[33] SAGE-CGCNN was trained with a batch size of 128 for up to 100 epochs. SAGE-CartNet was trained with a batch size of 128 for up to 300 epochs. SAGE-DenseGNN-Lite was implemented using TensorFlow/Keras through the kgcnn framework.[30] Models were trained with a batch size of 128 for up to 300 epochs. We use 0-based indexing in code; thus, middle_fusion_layers=2 corresponds to inserting SGMP at the third DenseGNN-Lite block. Text representations were obtained with MatSciBERT via the HuggingFace Transformers stack.[31,51] Data preprocessing and evaluation were performed using standard scientific Python packages (NumPy, Pandas, SciPy, and scikit-learn).[52-54] All experiments were conducted on an RTX 5090 GPU (32 GB).

## ASSOCIATED CONTENT

The code and full model and training configurations used in this work are available on GitHub at https://github.com/zhanggh1999/SAGE-Net.

### Supporting Information

Text semantics and keyword taxonomy; model architecture and optional modules; additional regression results; additional interpretability case studies; effect of text preprocessing; methods and training details.

### Author Contributions

Conceptualization was conducted by G.Z. and D.H. G.Z. performed the investigation, data curation, and wrote the original draft. Y.Y. and Y.J. contributed to validation. K.B.S, and D.O.S contributed to software development and resources. Project administration and funding acquisition was handled by D.H. and L.Z. All authors have read, revised, and agreed to the published version of the manuscript.

### Notes

The authors declare no competing financial interest.

## ACKNOWLEDGMENTS

We acknowledge the support by the National Key Research and Development Program of China (Grant No. 2022YFA1402500), the National Natural Science Foundation of China (Grants No. 12404268, 62125402, T2521005), and the Fundamental Research Funds for the Central Universities, JLU. Calculations were performed in part at the High-Performance Computing Center of Jilin University.